\documentclass[prl, twocolumn, preprintnumbers,amsmath,amssymb,superscriptaddress]{revtex4}
\usepackage{graphicx}
\usepackage{dcolumn}
\usepackage{bm}
\usepackage{subfigure}
\usepackage{multirow}
\usepackage{amsmath}
\usepackage{color,soul}
\usepackage[font=footnotesize, justification=raggedright,singlelinecheck=false]{caption}
\begin{document}
\title{Barrier-Controlled Non-Equilibrium  Criticality in Reactive Particle Systems }

\author{Qun-Li Lei}
\affiliation{School of Chemical and Biomedical Engineering, Nanyang Technological University, 62 Nanyang Drive, 637459, Singapore}

\author{Hao Hu}
\affiliation{School of Physics and Materials Science, Anhui University, Hefei 230601, China}

\author{Ran Ni}
\email{r.ni@ntu.edu.sg}
\affiliation{School of Chemical and Biomedical Engineering, Nanyang Technological University, 62 Nanyang Drive, 637459, Singapore}

\begin{abstract}
Non-equilibrium critical phenomena generally exist in many dynamic systems, like chemical reactions and some driven-dissipative {reactive} particle systems.  Here, by using computer simulation and theoretical analysis, we demonstrate the crucial role of the activation barrier on the criticality of {dynamic phase transitions in} a minimal reactive hard-sphere model.  We find that at zero thermal noise, with  increasing the activation barrier, the type of transition  changes from a continuous conserved directed percolation into a discontinuous dynamic transition by crossing a \emph{tricritical} point. A mean-field theory combined with field-simulation is proposed to explain this phenomenon. {The possibility of Ising-type criticality in the non-equilibrium system at finite thermal noise is also discussed.}
\end{abstract}
\maketitle

\section{Introduction}
Non-equilibrium critical phenomena~\cite{hin2000non,henkel2008non}  generically exist in various systems, e.g.,  chemical reactions~\cite{schlogl1972,claycomb2001}, epidemic spreads~\cite{pastor2015epidemic}, brain activities~\cite{munoz2018c}, colloidal and granular systems~\cite{pine2005chaos,corte2008random,regev2015,bak1987self},  turbulence and active fluids~\cite{ta2007directed,schaller2011frozen,shi2013,
partridge2019critical,sh2015a,hu2016no} etc. At the critical point, these systems exhibit scaling-invariance accompanied by diverging correlation lengths and time scales, which can be characterized by some generic critical exponents~\cite{henkel2008non}. The  directed percolation (DP) exemplified by the lattice contact model~\cite{harris1974c} is among the best-known universality classes of non-equilibrium phase transitions, yet deviations from the DP universality are also found in many dynamic systems~\cite{henkel2008non}. For examples, by introducing particle number conservation, a new universality of conserved direction percolation (C-DP) was found in the Manna model~\cite{manna1991two} and conserved lattice gas model~\cite{rossi2000}, while in the Schlogl's second model aiming to model the cooperative surface catalytic reaction, the dynamic transition becomes  first-order~\cite{schlogl1972}. {\color{black}  Nevertheless, lattice models are usually over-simplified and may neglect some key mechanisms in real dynamic critical systems}, e.g., chemical reactions, in which the kinetic energy, particle inertia, momentum conservation and some barrier-crossing processes play important roles.  Thus, it remains unclear whether there is any overlooked but fundamental mechanism controlling the non-equilibrium criticality in those real dynamic systems.

To this end, here we investigate a minimal reactive hard sphere model with an activation barrier, which captures the essential physics of many non-equilibrium particle systems. For examples, this model can be regarded as a simplified version of the Semenov hard sphere model for exothermic chemical reactions~\cite{chou1982,chou1984,baras1989,nowakowski2001s}. The Semenov model, which assumes a uniform temperature distribution and neglects the reactant consumption at ignition, is a classical theory of thermal explosion~\cite{semenov1928,semenov2013some}. Our model can also be viewed as a modification of  random-organization hard sphere model with the addition of an  activation barrier. The random-organization was originally proposed to model colloidal particle system under oscillatory shearing~\cite{pine2005chaos,corte2008random,pine2009prl,
f2011transverse,j2014geometric}, while it exhibits generic non-equilibrium critical behavior, which has connections with  the yielding of amorphous solids~\cite{regev2015,2017yielding}, the jamming transition~\cite{das2020unified,ness2020absorbing,milz2013,zhou2014random,nagasawa2019},  depinning transition~\cite{mangan2008reversible,reich2009random,okuma2011} and dynamic hyperuniform states~\cite{hexner2015,tjhung2015,weijs2015emergent,lei2019,lei2019h,ma2019h} etc. By using simulation and theory, we prove that the activation barrier controls the criticality or ``sharpness'' of the no-equilibrium phase transition in the system: at zero thermal noise, increasing the activation barrier changes the type of absorbing transition from a continuous C-DP to a discontinuous dynamic transition by crossing a \emph{tricritical} point. {We propose a mean-field theory combined with field-simulation to explain this phenomenon. The mean-field analysis also indicates that the non-equilibrium transition at finite thermal could be Ising-type}.

\section{Results}
\subsection{Model and Simulation}
The model we consider  consists of $N$  hard spheres  with the same mass $m$ and diameter $\sigma$.  Activations occur in pairwise collisions between two particles { $i, j$}, if their {relative distance approaches $\sigma$ and}  relative kinetic energy along the center-to-center direction surpasses the activation barrier $E_b$, i.e., 
{ 
\begin{eqnarray}
|\Delta {\mathbf r}_{i,j} | &=& \sigma  \\
\frac{1}{2}m \left(\Delta {v}^{\perp}_{i,j} \right)^2 &>& E_b
\end{eqnarray}
where $\Delta {\mathbf r}_{i,j}  = {\mathbf r}_{j}- {\mathbf r}_{i}$ is the relative distance between two particles and
\begin{eqnarray}
\Delta {v}^{\perp}_{i,j}  &=& \Delta {\mathbf v}_{i,j}\cdot \frac{\Delta {\mathbf r}_{i,j} } {\sigma}
\end{eqnarray}
is the relative velocity along the center-to-center direction with $\Delta {\mathbf v}_{i,j}  =  {\mathbf v}_{j}- {\mathbf v}_{i}$ the relative velocity.} During each activation, additional kinetic energy $\epsilon$ is released, with the momentum of the two particles conserved. { We assume a frictionless collision, thus the velocity change for particle $i,j$ during the collision  is 
\begin{eqnarray}
{\mathbf u}_{i}  &=& -{\mathbf u}_{j} =\frac{\Delta { v}^{\perp}_{i,j} - \sqrt{(\Delta { v}^{\perp}_{i,j})^2 + 4 \epsilon/m } }{2} \frac{\Delta {\mathbf r}_{i,j} } {\sigma}.
\end{eqnarray} }
Between two consecutive collisions, the motion of particles obeys the Langevin dynamics. For particle $i$, the equation of motion can be written as
\begin{eqnarray} \label{langevin_Eq}
m\frac{d {\mathbf v_i}(t)}{d t}=-\gamma {\mathbf v_i}(t)+\sqrt{2\gamma k_B T}\eta(t),
\end{eqnarray}
where $\gamma$ is the damping coefficient.  The second noise term is based on the fluctuation-dissipation theorem with $T$ the thermal temperature and  $\eta(t)$ the Gaussian white noise.   This model  is a direct generalization of the random-organizing hard-sphere model with the addition of an activation barrier and thermal noise{\color{black}, which can be realized by using active spinners with short-range repulsion}~\cite{lei2019h}.   It can also be seen as a simplified  version of  the Semenov reactive hard-sphere model for exothermic chemical reactions, in which non-consumable  reactants are confined by two parallel walls connected with thermal reservoir at a fixed temperature~\cite{chou1982,chou1984,baras1989,nowakowski2001s}, whereas the boundary thermalization is treated implicitly through a Langevin thermostat in our model.

We adopt the Langevin dynamic event-driven simulation to simulate this system. The evolution of the velocities and positions of particles, as well as the collisions between particles, are all deterministic in the  absence of thermal noise. The prediction of collisions and event-management are well documented in Ref.~\cite{lei2019h,rapaport2004art}. In the presence of the thermal noise, thermalization events are added to the event calendar after every time step $\Delta t =\tau_d/10 $, { where $\tau_d = m/\gamma$ is the typical dissipation time.} The velocity of particle $i$ in the case of no collision is updated according to \cite{scala2012event}
\begin{eqnarray}
v_{i,\alpha}(t + \Delta t) = e^{-\frac{\gamma}{m} \Delta t } v_{i,\alpha}(t) + \sqrt{v_{th}^2(1-e^{-2\frac{\gamma}{m}\Delta t }) } ~ \eta(\mathbf{r},t) \nonumber \\
\end{eqnarray}
where $\alpha=x,y,z$ and  $v_{th}^2 = dk_B T/m$. The dimensionality of the system we study is $d=2, 3$ with periodic boundary conditions in all directions. The reduced particle density of the system is defined as $\tilde{\rho}=N\sigma^d/V$ with $V=L^d$ the volume of the system and $L$ the box length. The typical excitation speed is defined as $v_0=\sqrt{\epsilon/m}$, and the time unit of the system is set as $\tau_0=\sigma/v_0$.

\begin{figure}[!bhtp] 
	\resizebox{83mm}{!}{\includegraphics[trim=0.0in 0.0in 0.0in 0.0in]{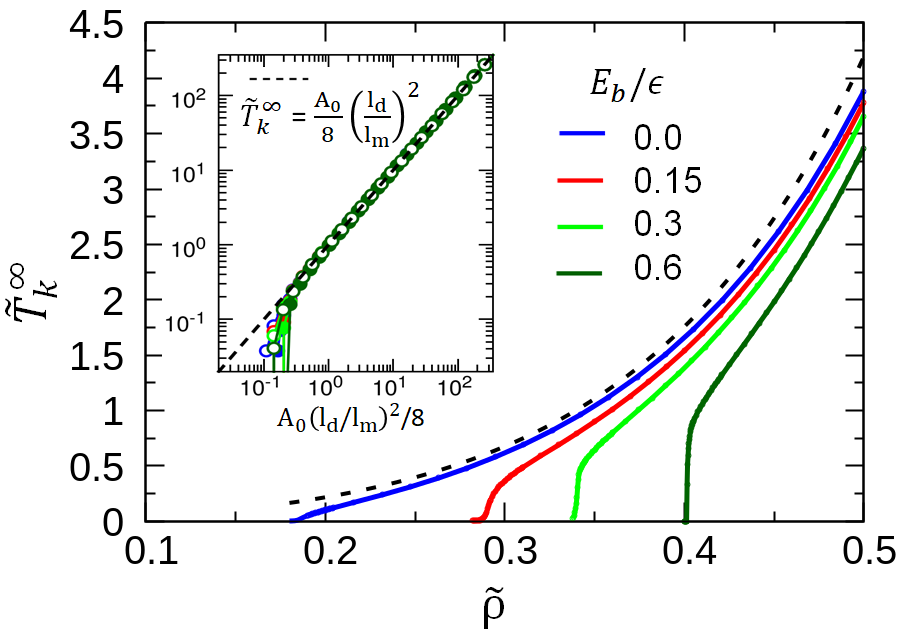} }
\caption{steady-state kinetic temperature $\tilde{T}_k^{\infty}$ as a function of  density for 2D system with different $E_b$. Inset:  $\tilde{T}_k^{\infty}$ as a function of  $l_d$ ($\tilde{\rho}=0.1$, {solid symbols}) and $l_m$ ($l_d=160\sigma$, {open symbols}).  The dashed lines are the theoretical predictions with $A_0=1.56$. }
\label{Fig1}
\end{figure}

\begin{figure*}[!htb] 
	\resizebox{177mm}{!}{\includegraphics[trim=0.0in 0.0in 0.0in 0.0in]{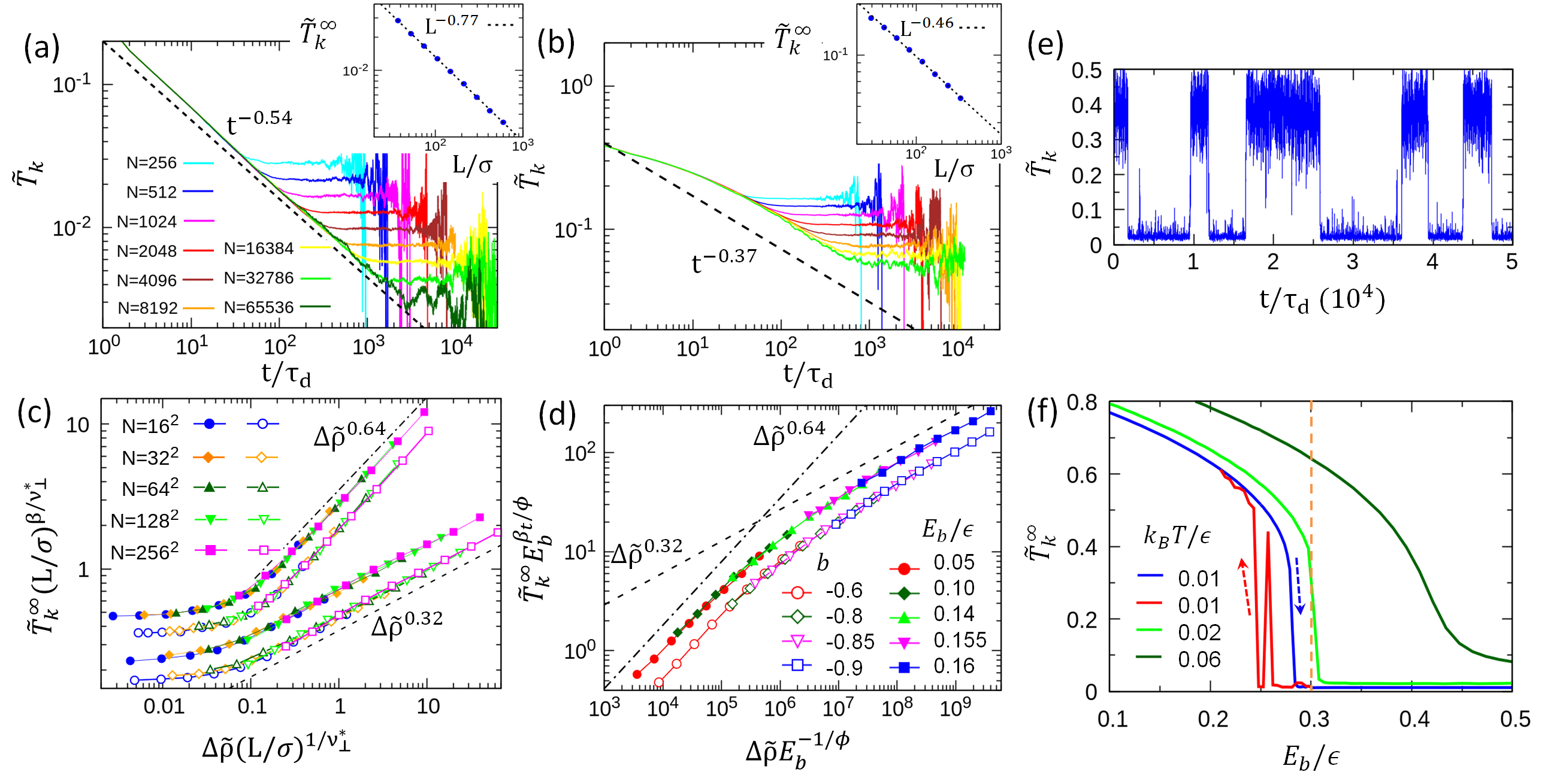} }
\caption{ Barrier-controlled critical behaviors in 2D systems. (a, b) $\tilde{T}_k$ as a function of time  for different system sizes with inset showing saturated value as a function of system size { at zero thermal noise}. {(a)}  $ E_b = 0,~ \tilde{\rho}=0.18471$;  ({b})  $E_b = 0.165 \epsilon$, $\tilde{\rho}=0.29615$.  ({c})  Collapse of $\tilde{T}_k (\Delta \tilde{\rho} )$ according to Eq.~(\ref{scaling1}) for different system sizes { at zero thermal noise}. Solid symbols are for reactive hard-sphere systems with  $E_b=0. 0$ (upper), $0.165\epsilon$ (lower), respectively. The open symbols are  obtained in field simulations  of Eq.~(\ref{thermophoresis1}-\ref{thermophoresis2}) with $b=1$ (upper),  -0.92 (lower), respectively.  ({d})  Crossover scaling analysis of the tricritical point based on Eq~(\ref{cross_over_scaling}) { at zero thermal noise}. ({e})  Demonstration of bistability at $ E_b = 0.3\epsilon,~ \tilde{\rho}=0.3290$ with thermal noise $k_BT=0.02\epsilon$  for a system with  $N=1024$. ({f})  $\tilde{T}_k$ as a function of reaction barrier $E_b$ at different thermal temperature $T$ for 2D hard sphere systems  at density  $\tilde{\rho}=0.3290$ and $N=1024$. Hysteresis loop is found in system at $k_B T=0.01\epsilon$ starting from different initial states as indicated by the dashed arrows. The yellow line indicates the density point at which the system exhibits clear bistability in {(e)}. }
\label{Fig2}
\end{figure*}

\subsection{Critical phenomena}
We first study the 2D system at zero thermal noise $T=0$.  In this case, as shown in Ref~\cite{lei2019h}, there are two important characteristic lengths in the system, i.e., the mean free path $l_m$ and the dissipation length $l_d ={\sqrt{ m \epsilon } }/{{\gamma} }$.  The later is the typical distance that an isolated particle can travel after being activated by collision. The system  undergoes a dynamic absorbing-active phase transition at $l_d \simeq l_m$~\cite{lei2019h}. Here, the absorbing state corresponds to the zero kinetic energy state {or zero kinetic temperature $T_k=0$,} while the active state has a finite kinetic energy {($T_k>0$)}.  { The kinetic temperature of the system $T_k$ is different from $T$, which reflects the strength of thermal noise.} Thus, the reduced kinetic temperature of the system can be chosen as the order parameter of the system, i.e.,
\begin{eqnarray} \label{T_k}
\tilde{T}_k(t) =\frac{d k_B T_{k}(t)}{\epsilon} ~~\mathrm{and} ~~ k_B T_{k}(t) = \frac{m\langle \overline{{v}^2}(t) \rangle  }{d}.
\end{eqnarray}
Here, $\overline{v^2} $ is the average square of speed of all particles and  $\langle \cdot \rangle$ calculates the ensemble average of active (surviving) trials~\cite{rossi2000}. In Fig.~\ref{Fig1}, we plot the steady-state kinetic temperature $\tilde{T}_k^{\infty} \equiv \tilde{T}_k (t \rightarrow \infty)$ as a function of density for systems with different activation barriers. With increasing $E_b$, the transition shifts to the higher density regime and becomes sharper, similar to the explosive percolation~\cite{achlioptas2009explosive}. Same phenomenon is also found in 3D systems as shown Fig.~S1 in Ref.~\cite{Supplemental2}.  

To determine the type of the transitions, we  assume there is a scaling-invariant critical point $\tilde{\rho}_c$  in the system and perform finite-size analysis to determine $\tilde{\rho}_c$ and the corresponding critical exponents. For systems near the critical point, $\tilde{T}_k^{\infty}(\Delta \tilde{\rho}, L)$ satisfies the  scaling relationship~\cite{henkel2008non,rossi2000},
\begin{eqnarray} \label{scaling1}
\tilde{T}^{\infty}_k(\Delta \tilde{\rho}, L) = L^{-\beta/\nu^*_{\perp}} \mathcal{G}\left(L^{1/\nu^*_{\perp}} \Delta \tilde{\rho}\right),
\end{eqnarray}
where $\Delta \tilde{\rho} = \tilde{\rho}- \tilde{\rho}_c$ and $\mathcal{G}(\cdot)$ is the scaling function. $\nu^*_{\perp}$  normally equals $\nu_{\perp}$ in the scaling relationship of spatial correlation length $\xi_{\perp} \sim |\Delta \tilde{\rho}|^{-\nu_{\perp}}$~\cite{lei2019h}. For systems at the critical point ($\Delta \tilde{\rho}=0$), starting from random initial configurations, $\tilde{T}_k(t)$  follows a power-law decay $t^{-\alpha}$ before reaching the saturated value satisfying  $\tilde{T}^{\infty}_k \sim L^{-\beta/\nu^*_{\perp}}$. In all our simulations, we use natural homogeneous initial configurations~\cite{basu2012}, which have been shown able to avoid the anomalous undershooting and lead to a more self-consistent  $\alpha$ exponent compared with complete random initial configurations~\cite{basu2012,lee2013comment}. In Fig.~\ref{Fig2}a, we show the decay of $\tilde{T}_k(t) $ for a 2D system of different  size $L\propto {N}^{1/d}$ for $E_b=0$ and $l_d=2\sigma$. The inset plots $\tilde{T}^{\infty}_k$ as a function of $L$. Both figures indicate the scaling-invariance  and we obtain the corresponding critical exponents $\alpha=0.54$ and $\beta/\nu^*_{\perp}=-0.77$. In Fig.~\ref{Fig2}c, we plot the collapse of $\tilde{T}^{\infty}_k(\Delta \tilde{\rho}, L)$ for different  $L$ (solid symbols) based on Eq.~(\ref{scaling1}) with $\beta=0.64$ and $\nu^*_{\perp}=0.84$, which is consistent with the  obtained  $\beta/\nu^*_{\perp}$.  

\begin{figure*}[htbp] 
	\resizebox{135mm}{!}{\includegraphics[trim=0.0in 0.0in 0.0in 0.0in]{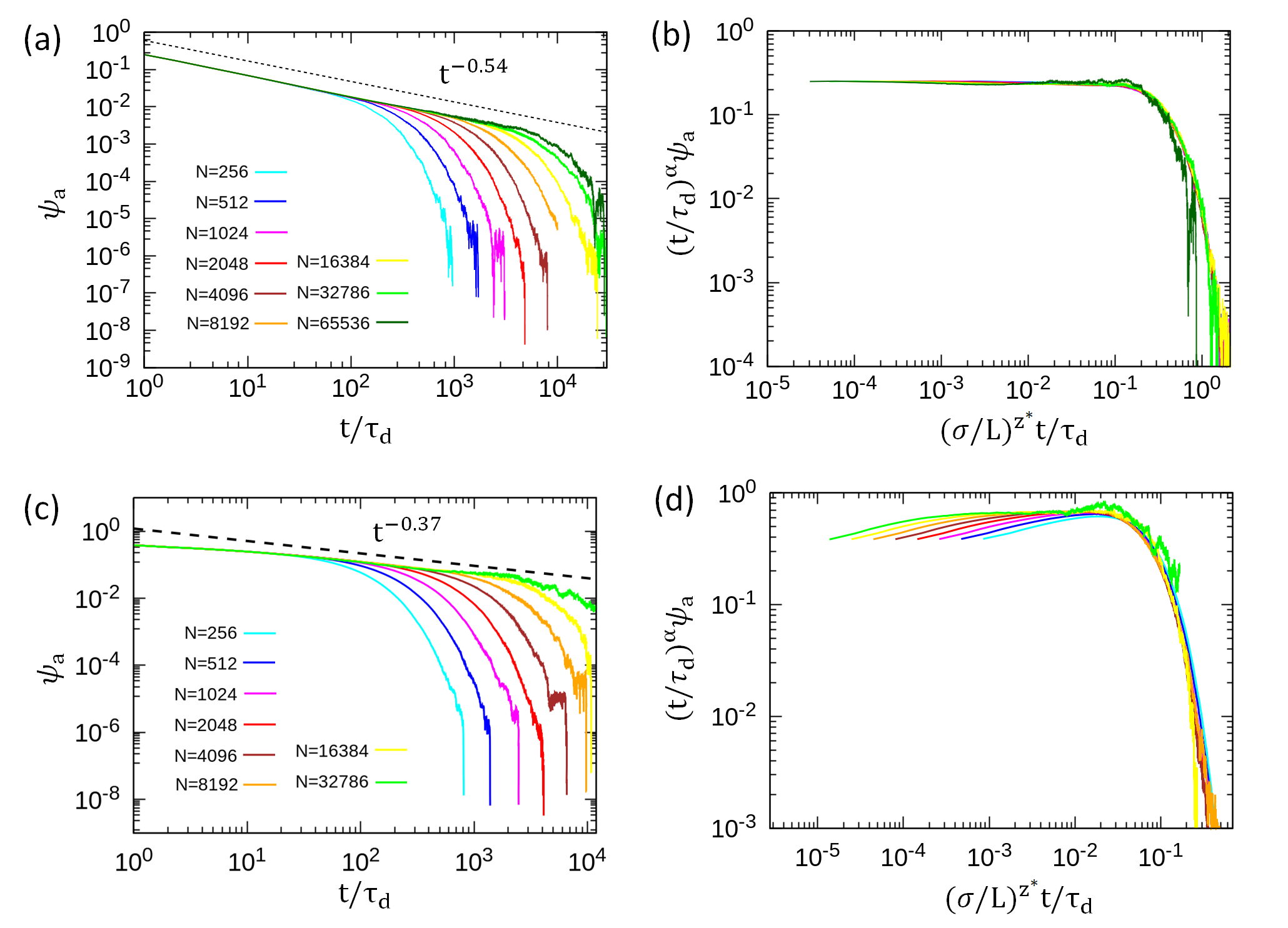} }
\caption{ Overall activity $\psi_a(t)$ for different system sizes in 2D hard sphere systems at $ E_b = 0,~ \tilde{\rho}=0.18471$ .  (b) Collapse of activity $\psi_a(t)$ based on Eq.(\ref{finite_size_2}) with  $\alpha=0.54$, $z^*=1.50$.  (c) Overall activity $\psi_a(t)$ for different system sizes in 2D  at  tricritical point $E_b = 0.165 \epsilon$, $\tilde{\rho}=0.29615$.  (d) Collapse of activity $\psi_a(t)$ with  $\alpha=0.37$, $z^*=1.68$.}
\label{Fig3}
\end{figure*}

At the critical state, one can also define the overall activity $\psi_a$, which has the same definition as $\tilde{T}_{k}$ except that the data are averaged over both surviving and non-surviving trials~\cite{rossi2000}. $\psi_a$  obeys another finite-size scaling
\begin{eqnarray}  \label{finite_size_2}
\psi_a(t,L) = t^{-\alpha} \mathcal{F}\left(tL^{-z^*}\right)
\end{eqnarray}
with $\mathcal{F}(\cdot)$ the scaling function. Here $z^*=\nu_{\parallel}/\nu^*_{\perp}$ is the dynamic exponent with $\nu_{\parallel}$ originating from the scaling relationship of temporal correlation length,  $\xi_{\parallel} \sim |\Delta \tilde{\rho}|^{-\nu_{\parallel}}$. In Fig.~\ref{Fig3}a,b, by using this finite-size scaling method, we obtain $z=1.50$.  

All the critical exponents we obtain at $E_b=0$ and {zero thermal noise} for both 2D and 3D systems are summarized in Table I, which indicates that the criticality of this  phase transition belongs to C-DP or Manna class~\cite{manna1991two,lei2019h}. Nevertheless, with a similar analysis for 2D systems of $E_b=0.165\epsilon$ in Fig.~\ref{Fig2}b,d and Fig.~\ref{Fig3}c,d, we obtain critical exponents distinct from C-DP (Table I). Same finding applies to 3D systems as shown in Fig.~S2-S3 in Ref.~\cite{Supplemental2}.  As shown later, these unreported exponents appear because the system is at a \emph{tricritical} point, which separates the continuous and  discontinuous transitions. Indeed, by further increasing $E_b$ to $0.3\epsilon$, we find the finite-size scaling breaks down and the 2D system exhibits the bistability {if turning on the thermal noise to} $k_BT=0.02\epsilon$ (Fig.~\ref{Fig2}e). In Fig.~\ref{Fig2}f, we further show the appearance of cusp bifurcation and hysteresis loop with increasing the activation barrier  { at finite thermal noise}. Therefore, the system undergoes a discontinuous transition at large activation barrier even with a finite thermal noise.

To determine the tricritcal point {at the zero thermal noise}, we  employ a crossover scaling analysis~\cite{lubeck2006tricritical,janssen2004g,araujo2011tricritical}. Supposing $E_{b,c}$ is the tricritical energy barrier and letting $\Delta E_b  = E_b - E_{b,c}$,  the crossover scaling can be written as
\begin{eqnarray}\label{cross_over_scaling}
\tilde{T}_{k}(\Delta \tilde{\rho},\Delta E_b) = \Delta E_b^{-\beta_t/\phi}  \mathcal{G}_t( \Delta \tilde{\rho}  \Delta E_b^{-1/\phi}),
\end{eqnarray}
given $\Delta \tilde{\rho} \ll \Delta E_b/\epsilon$~\cite{lubeck2006tricritical}. Here, the scaling function $  \mathcal{G}_t(x)\sim x^\beta$ when $x\ll 1$ and  $ \mathcal{G}_t(x)\sim x^{\beta_t}$ when $x\gg 1$  with ${\beta_t}$ the tricritical exponent.  In Fig.~\ref{Fig2}d, we show the collapse of $ \tilde{T}_{k}(\Delta \tilde{\rho})$  for 2D systems with different $E_b$ (solid symbols), which leads to  $E_{b,c}=0.165\epsilon$ and  $\phi=0.32$. Similar collapsing can be made to determine the tricritical point in 3D systems as shown in Fig~S2 in Ref.~\cite{Supplemental2}.

\begin{table}[!htb] 
\begin{tabular}{c c c c c c}
\multicolumn{6}{c}{2D}  \\
\hline \hline \\[-2.0ex]
 ~~& ~~~~$\beta_{(t)}$~~~~ & ~~~~$\alpha$~~~~ & ~~~~$\nu_{\perp}^*$~~~~ & ~~~~$z^*$~~~~& ~~~~$\phi$~~~~  \\[0.5ex]
\hline 
\\[-1.5ex]

 C-DP~~ & 0.64(1)  & 0.52(1) & 0.80(2)  & 1.53(2) &   \\ [0.5ex]

$E_b=0$               & 0.64(2) & 0.54(2) & 0.84(3) & 1.50(3)  &  \\ [0.5ex]

$b=1$                  & 0.65(2) & 0.53(2) & 0.84(3) & 1.51(2)  & \\ [0.5ex]

$E_b=E_{b,c}$  & 0.32(4) & 0.37(5) & 0.68(4) & 1.68(5)  & 0.32(5) \\ [0.5ex]

$b=b_c$          & 0.33(3) & 0.36(4) & 0.70(3) & 1.65(5)  & 0.34(4) \\ [0.5ex]

\hline \\

\multicolumn{6}{c}{3D}  \\
\hline \hline \\[-2.0ex]
 ~~& ~~~~$\beta_{(t)}$~~~~ & ~~~~$\alpha$~~~~ & ~~~~$\nu_{\perp}^*$~~~~ & ~~~~$z^*$~~~~& ~~~~$\phi$~~~~  \\[0.5ex]
\hline 
\\[-1.5ex]

 C-DP~~ & 0.84(1)  & 0.75(2) & 0.59(1)  & 1.82(2) &   \\ [0.5ex]

$E_b=0$  & 0.85(3) & 0.86(2) & 0.60(2) & 1.79(4)  &  \\ [0.5ex]

$b=1$    & 0.86(3) & 0.86(2) & 0.59(3) & 1.78(3)  & \\ [0.5ex]

$E_b=E_{b,c}$  & 0.48(3) & 0.52(4) & 1.0(4) & 1.95(6)  & 0.50(5) \\ [0.5ex]

$b=b_c$          & 0.50(2) & 0.51(3) & 0.95(5) & 2.0(4)  & 0.50(4) \\ [0.5ex]

Mean field~~      & 1/2    & 1/2    & 1      & 2        & 1/2  \\ [0.5ex]

\hline

\end{tabular}
\caption{ { Critical exponents  in  2D and 3D systems}.  Data for C-DP is from \cite{henkel2008non,lee2013comment}. Different values of $E_b$ and $b$ are for reactive hard-sphere model and field simulations  of Eq.~(\ref{thermophoresis1}-\ref{thermophoresis2}), respectively. Tricritical points for 2D system are at $E_{b,c}=0.165(5)\epsilon$ and $b_{c} = -0.92(3)$.  Tricritical points for 3D system  at $E_{b,c}=0.070(3)\epsilon$ and $b_{c} = -0.35(2)$. Note that in 3D case, the $\alpha$ is  larger than the reference value because we use  natural homogeneous initial configurations in our simulations~\cite{basu2012,lee2013comment}.} \label{Tab_1}
\end{table}

\subsection{General theoretical analysis}
{\color{black} To understand the complex phase behaviour above, we first analyse the variables, intrinsic symmetries, conservation laws, and  forms of fluctuations in the system: i) generally, the system is described by three hydrodynamic variables, i.e., the density field, the velocity field and the energy field. However, since the velocity field is damped, it is no longer an independent  variable in the hydrodynamic limit; ii) the system is spatially isotropic, so the  gradient terms in the dynamic equation must be even; iii) the system conserves the particle number, but does not conserve the momentum and energy; iv) the fluctuations come from two parts: one is from the thermal excitation whose variance is proportional to the thermal temperature $T$, the other is from collision activations with the variance proportional to the kinetic temperature $T_{k}$. As a result, the dynamic equations for the left two independent variables, local energy (kinetic temperature) field $\tilde{T}_k^l (\mathbf{r},t)$ and local density field $\tilde{\rho}_l (\mathbf{r},t)$ of the system should have the form 
 \begin{eqnarray}
\frac{\partial  \tilde{T}_k^l  }{\partial {t}} &=&  \mu \nabla^2 \tilde{T}_k^l + F( \tilde{T}_k^l, \tilde{\rho}_l, T) + \Lambda \sqrt{ \tilde{T}_k } \eta(\mathbf{r},t)~~~~~~~ \label{general_equation_1} \\
\frac{\partial}{\partial t}  \tilde{\rho}_l &=& \nabla \cdot {\mathbf J(\tilde{T}_k^l, \tilde{\rho}_l) } + ~ \nabla \cdot {\boldsymbol \xi}(\mathbf{r},t). \label{general_equation_2}
\end{eqnarray}
Here, the second equation is a result of particle number conservation. Function $F$ and current $\mathbf J$ can be arbitrary functions of $\tilde{\rho}$, $\tilde{T}_k$, $T$ and their spatial derivatives. $\Lambda$ is the strength of Gaussian white noise $\eta(\mathbf{r},t)$. Noise ${\boldsymbol \xi}(\mathbf{r},t)$  satisfying $\langle \xi_i(\mathbf{r},t) \xi_j(\mathbf{r}',t') \rangle = 2 \Gamma \tilde{\rho} k_B T \delta_{ij} \delta(\mathbf{r}-\mathbf{r}') \delta(t-t')$. Near the critical point, only the leading terms in $F$ and current $\mathbf J$ are important.  Without thermal excitation ${\boldsymbol \xi}(\mathbf{r},t)=0$, the mass transport can only happen when both $\tilde{T}_k$ and $\tilde{\rho}$ are non-zero, thus the leading terms in $\mathbf J$ would be proportional to $\tilde{T}_k \nabla \tilde{\rho} $ and $ \tilde{\rho} \nabla  \tilde{T}_k$. 
}

\begin{figure}[!bhtp] 
	\resizebox{85mm}{!}{\includegraphics[trim=0in 0.0in 0.0in 0.0in]{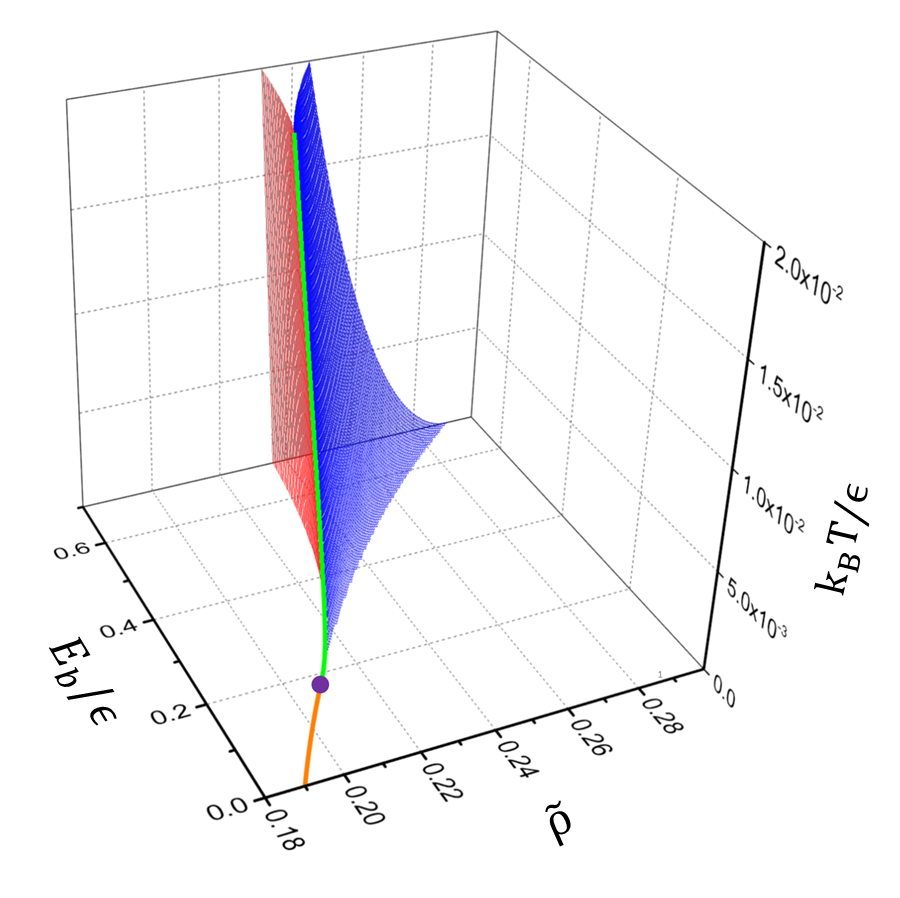} }
\caption{ Mean-field phase diagram. The orange line and green line represent of the  C-DP and Ising-type critical points of the system, respectively. The purple point is the tricritical point. The bistability region is enclosed by the blue and red surfaces. The parameters used is $A=0.095$, $B=1/2$ and $\tau_d/\tau_0=10$. }
\label{Fig4}
\end{figure}

\subsection{Mean-field theory}
{\color{black}Now we obtain the expression for $F(\tilde{T}_k, \tilde{\rho}, T)$ from the kinetic point of view.} At the mean-field level, the activation driving power $W_{driv}$ and  the  dissipation power $W_{disp}$  per particle are ~\cite{lei2019h}
 \begin{eqnarray}\label{disp_driv}
 W_{driv}= f_a \epsilon,~~~~  W_{disp} = \overline{v^2} \gamma 
\end{eqnarray}
respectively.  Here, $f_a$ is the average activating collision frequency per  particle which can be written as
 \begin{eqnarray}\label{f_a}
f_a =  x_a \overline{v}_a/(2l_r) ,
\end{eqnarray}
 with  $x_a$  the  fraction of activated particles and  $ \overline{v}_a $ is the average speed of activated particles.   $l_r$ is the mean free path of activating collisions. At $E_b=0$,  $l_r$  equals the mean free path of the system $l_m$ since every collision induces an activation, while for $E_b>0$, one can expect $l_r > l_m$ as discussed latter.  Moreover, the  driving power of thermal noise per particle $W_{driv}^{therm}$ can be approximated by the equilibrium dissipation power $ W_{disp}^{therm}=\gamma d k_B T/m$.  Thus, {\color{black} by neglecting the fluctuation,} the mean-field dynamic equation for $\tilde{T}_k$ can be written as
 \begin{eqnarray}\label{dy_eq}
{\epsilon}\frac{\partial  \tilde{T}_k }{\partial t} &=& W_{driv} -W_{disp} +W_{driv}^{therm} \nonumber  \\
&=& \frac{x_a \tilde{\rho}_r \overline{v}_a\epsilon}{2 \sigma} - \gamma \overline{v^2}   + \frac{\gamma d k_B T}{m}.
\end{eqnarray}
Here we use the low density approximation for the mean free path of activating collision $l_r\simeq \sigma/ \tilde{\rho}_r$ with  $\tilde{\rho}_r$ the density of effective reactant, i.e., the average  density of particle  that can be activated under $\tilde{T}_k$.   For systems far from the critical point ($l_d\gg l_m$) the average  kinetic energy would be much larger than the activation barrier ($\tilde{T}_k \gg  E_b/\epsilon$), thus we  have $\tilde{\rho}_r \simeq  \tilde{\rho}$, $x_a\simeq 1$  and $ \overline{v}_a\simeq  \overline{v} $ with $ \overline{v} $ the average speed of all particles.   Based on the approximation $\overline{v}^2 \simeq \overline{v^2}=dk_B T_k/m$,  at zero thermal noise $T=0$,  the  steady-state kinetic temperature $\tilde{T}_k^{\infty}$ satisfying $\frac{\partial  \tilde{T}_k }{\partial t}=0$  can be obtained as ~\cite{lei2019h}
 \begin{eqnarray}
\tilde{T}_k^{\infty} \simeq  \frac{1}{4d}\left(\frac{l_d}{l_m}\right)^2,
 \end{eqnarray}
  which does not depend on $E_b$. This is verified in simulations as shown in the inset of Fig.~\ref{Fig1}.  Nevertheless, for systems close to the critical point, there is strong dynamic heterogeneity, i.e.,  most particles are  immobile and only a  small fraction of particles are activated with typical excitation speed $\overline{v}_a \simeq v_0$.  Apparently, $\overline{v}_a$ is a convex instead of concave function of $x_a$, since  increasing $x_a$  also increases the collision between activated particles, further raising $\overline{v}_a$.  Thus,  as a first-order approximation, we  have
\begin{eqnarray}
\overline{v}_a &\simeq &  (1 + A x_a)v_0  \\
\tilde{T}_k  &\simeq&  x_a m\overline{v}_a^2 /\epsilon  \simeq  x_a + 2A x_a^2
\end{eqnarray}
with the coefficient $A>0$ indicating the convexness.  Reversely, we can rewrite $x_a$ as a function of order parameter $\tilde{T}_k$:
\begin{eqnarray}
x_a \simeq  \tilde{T}_k  - 2A \tilde{T}_k^2.
\end{eqnarray}
Moreover, for systems  with $E_b>0$, one can expect that there would be a fraction of collisions between active and inactive particles that do not induce activation. This effect leads to the  elongation of the mean free path of activating collisions, or the decrease of the effective reactant density. Therefore, as a first-order approximation, we can have
\begin{eqnarray}
\tilde{\rho}_r/  \tilde{\rho} = 1- B  (1-x_a) \tilde{E}_b
\end{eqnarray}
with the coefficient $B>0$. Finally, by keeping only the first three leading terms,  Eq.~(\ref{dy_eq})  can be written as 
\begin{eqnarray}\label{MF_dy_eq}
\frac{\partial \tilde{T}_{k}}{\partial  \tilde{t} } = a \tilde{T}_{k} -b \tilde{T}_{k}^2 -c \tilde{T}_{k}^3 + h
\end{eqnarray}
with  $\tilde{t}=t/\tau_0$ and 
\begin{eqnarray}\label{abch}
a &=& \frac{ \tilde{\rho}}{2} (1-B \tilde{E}_b) -\frac{{\tau}_0}{{\tau}_d } \label{coeff_a}    \\
b &=&  -\frac{ \tilde{\rho}}{2}  \left[(1+A)B \tilde{E}_b-A\right]  \\
c &=&  \frac{ \tilde{\rho}}{2} \left[4 A^2+(3A-4A^2)B \tilde{E}_b\right]  \label{coeff_c}. \\
h &=&  {\gamma  \tau_0  d k_B T}/(m \epsilon)
\end{eqnarray}
The solution of this cubic mean-field dynamic equation is given in the Appendix A. The corresponding phase diagram  is summarized in Fig.~\ref{Fig4}. At zero  thermal noise ($T=0$) and $E_b<E_{b,c}$~(or $b>0$),  Eq.~(\ref{MF_dy_eq}) has only one fixed point at $a=0$ and the system undergoes a continuous phase transition with the mean-field critical exponent $\beta_{\rm MF}=1$ and upper critical dimension $d_c=4$. The transition points are shown as an orange curve in Fig.~\ref{Fig4}.  In contrast,  when $E_b$ increases above $E_{b,c}$ (or $b<0$),  Eq.~(\ref{MF_dy_eq}) has two fixed points, which corresponds to a bistability or a  discontinuous dynamic phase transition. Importantly, $E_b = E_{b,c}$ (or $b=0$) is the tricritical point with  $\beta_{\rm MF}=1/2$ and $d_c=3$, which separates the continuous and discontinuous phase transitions  (purple point in Fig.~\ref{Fig4}). In the presence of  thermal noise ($T>0$), the continuous phase transition is {smeared} out accompanied by the shrink of  bistability region (enclosed by red and blue surfaces).  This general picture  is insensitive to the specific choice of  coefficients $A, B$ and agrees qualitatively with  the simulation results.

\begin{figure*}[htbp] 
	\resizebox{180mm}{!}{\includegraphics[trim=0.0in 0.0in 0.0in 0.0in]{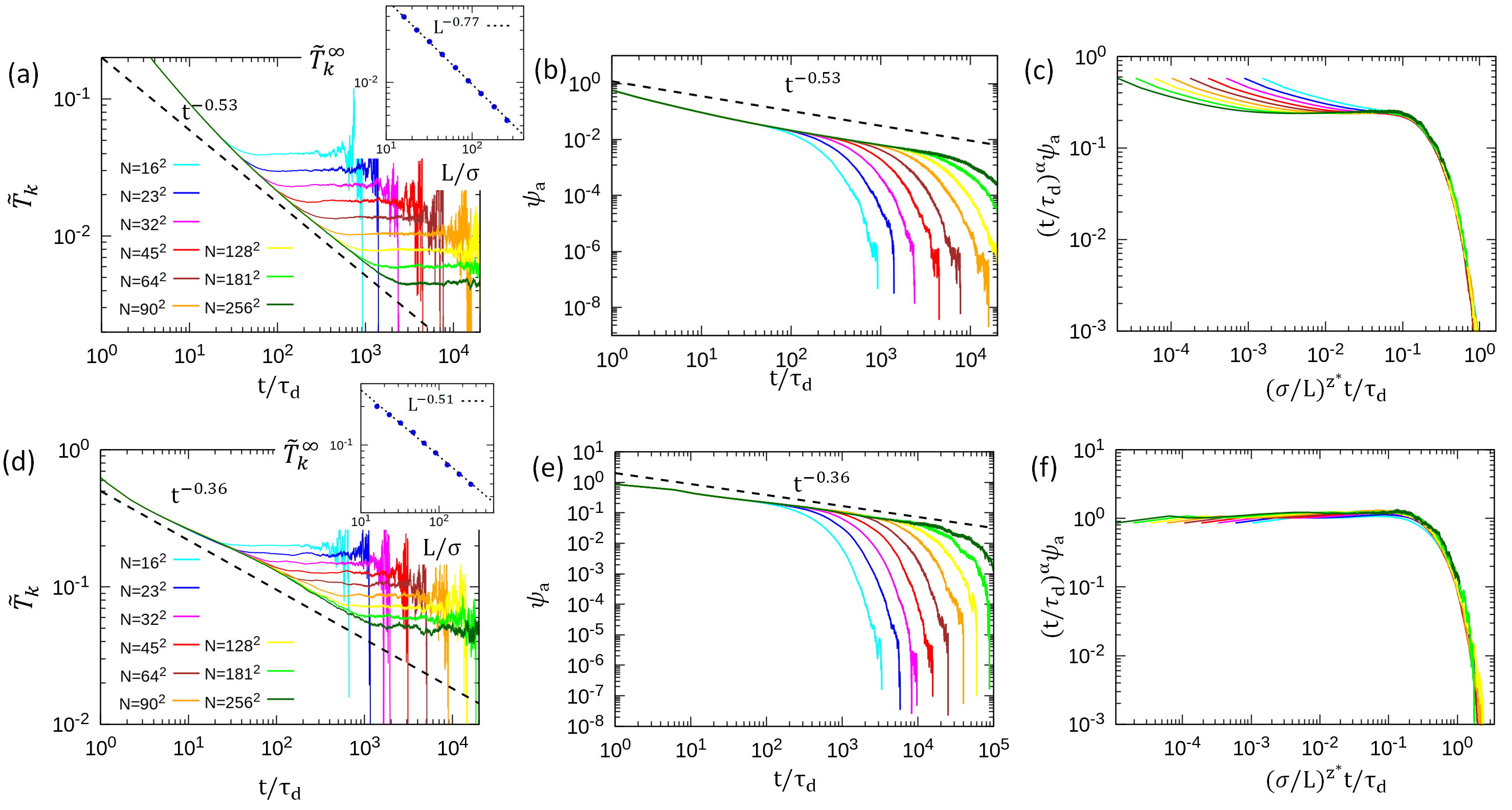} }
\caption{ Results for for 2D field simulations. (a) $\tilde{T}_k$ as a function of $t$ for different system sizes  with  $ b = 1$, $\tilde{\rho}= 1.3722$. (inset): saturated value as a function of system size $L$. (b) Overall activity $\psi_a(t)$ for different system sizes.  (c) Collapse of  $\psi_a(t)$ with  $\alpha=0.53$, $z^*=1.51$. (d) $\tilde{T}_k$ as a function of $t$ for different system sizes for 2D field simulation at tricritical point  $ b = -0.92$, $\tilde{\rho}= 0.8316$.  (e) Overall activity $\psi_a(t)$ for different system sizes at $ b = -0.92$.  (f) Collapse of  $\psi_a(t)$   with  $\alpha=0.36$, $z^*=1.65$.}
\label{Fig5}
\end{figure*}

\subsection{ Reggeon-field simulations at zero thermal noise}
{\color{black} Next we further go beyond the mean-field level and recover spatial-temporal fluctuations in our theory. At $T=0$, the density of the system is homogeneous. The only leading term in current $\mathbf J$ of Eq.~(\ref{general_equation_1}) is $\nabla^2 \tilde{T}_k^l$. Thus  Eq.~(\ref{general_equation_1}-\ref{general_equation_2}) can be rewritten as,}
 \begin{eqnarray}
\frac{\partial  \tilde{T}_k^l  }{\partial \tilde{t}} &=&  \mu \nabla^2\tilde{T}_k^l  + a \tilde{T}_k^l  - b\tilde{T}_k^{l2}  - c \tilde{T}_k^{l3}    + \Lambda \sqrt{ \tilde{T}_k }  \eta(\mathbf{r},t)~~~~~~~ \label{thermophoresis1} \\
\frac{\partial}{\partial t}  \tilde{\rho}_l &=& {D} \nabla^2 \tilde{T}_k^l, \label{thermophoresis2}
\end{eqnarray}
where, $a$, $b$ and $c$ are functions of  $\tilde{\rho}_l (\mathbf{r},t)$ based on Eq.~(\ref{coeff_a}-\ref{coeff_c}).  The second dynamic equation Eq.~(\ref{thermophoresis2}) now becomes  the thermophoreis diffusion equation of the particles. Eq.~(\ref{thermophoresis1}-\ref{thermophoresis2}), in fact, share the same formula as in the Reggeon-field theory of C-DP~\cite{di2016self}. Since the criticality of  dynamic equations does not depend on specific  choice of parameters, the critical exponents for  Eq.~(\ref{thermophoresis1}-\ref{thermophoresis2})  should be the same as the standard Reggeon field theory in which $b$ and $\tilde{\rho}$ are the controlling parameters with  $a=a'\tilde{\rho}_l (\mathbf{r},t)-a_0$. Other parameters are simply set as $a_0=1$, $a'=1$, $c=1$, $\lambda=1$, $\mu=D=1$, $\Lambda=1$~\cite{di2016self}. Since the noise term is multiplicative, we adopt a numerical method based on Fokker-Planck equation~\cite{dornic2005integration} to integrate the standard Reggeon fields equations on square or cubic lattices. This method has been used to obtain the accurate critical exponents for both directed percolation and C-DP~\cite{dornic2005integration}, but not for the tricritical points.  In Fig.~\ref{Fig2}c,d, the open symbols show the results from the field  simulations, which behave essentially the same as the reactive hard-sphere system. In Fig.~\ref{Fig5}, we do the same finite size scaling analysis for 2D field systems to obtain the critical exponents at $b=1$ and $b=-0.92$ (tricritical point). Similar analysis for 3D field systems can be found in Fig.~S4 in Ref.~\cite{Supplemental2}. The obtained (tri)critical exponents are also given in Table I, which are consistent with those from the hard-sphere systems. Moreover, the upper critical dimension $d_c=3$ at tricriticality is also confirmed both in 3D hard-sphere systems and field simulations (see Table I).  This quantitative agreement further proves the validity of our theory and concludes that the critical behavior of the reactive hard-sphere model at the edge of continuous absorbing phase transition is the same as the tricritical C-DP.

\subsection{Criticality at finite thermal noise}
Lastly, our theory  predicts  another  critical line (the green line in Fig.~\ref{Fig4})  in the  finite thermal temperature regime ($T>0$),  which at mean-field level satisfies $[a_c, b_c, \tilde{T}_{k,c}]=\left[-3\left(\frac{h}{c}\right)^{2/3},  -3\left(\frac{h}{c}\right)^{1/3},  \left(\frac{h}{c}\right)^{1/3} \right ]$.  Around the critical point $[a_c,~  b_c]$, we can find a  direction in the $(a,~b)$ parameter space, i.e.,  $(  \tilde{T}_{k,c}, ~1)$,  along which Eq.~(\ref{MF_dy_eq})  can be rewritten as 
\begin{eqnarray}\label{MF_pertur3}
\frac{\partial \Delta \tilde{T}_{k}  }{\partial \tilde{t}} &\simeq & -c \Delta \tilde{T}_{k} ( \Delta \tilde{T}_{k}^2 +  {  \tilde{T}_{k,c}\Delta b/c}), ~~~~~~
\end{eqnarray}
with $ \Delta \tilde{T}_{k} = \tilde{T}_{k} - \tilde{T}_{k,c}$ and $ \Delta b =b - b_{c}$ (see Appendix A). Therefore, for $\Delta b <0$, the system  has two stable fix points $
\Delta \tilde{T}_{k} \simeq  \pm \sqrt{-\Delta b \tilde{T}_{k,c}/c }$, which leads to $\beta_{\rm MF}=1/2$.   Eq.~(\ref{MF_pertur3}) also indicates that for $T>0$, the system has an emergent   ${\mathbb Z}_2$ symmetry under the transformation ($\Delta \tilde{T}_{k} \rightarrow - \Delta \tilde{T}_{k} $). Thus, { this mean-field analysis suggests} that phase transition for $T>0$ could be a Ising-type. { A recent work by Korchinski et al.~\cite{kor2019} has shown that in the presence of thermal noise, the broken symmetry between space and time in DP can be partially restored, which gives rise to undirected-percolation-like critical behaviors.} In fact, systems described by the dynamic equations for a non-conserved field  coupled to a conserved field i.e., Eq~(\ref{general_equation_1}-\ref{general_equation_2}), is categorised as dynamic model C according to Ref.~\cite{hohenberg1977}. { Nevertheless,} the relationship between the critical behaviors of non-equilibrium and the Ising universality has long been an active discussion topic in areas from chemical reactions~\cite{dewel1977,brachet1981,dewel1981,grassberger1981p,tome1993ziff,liu2004crossover}, to dynamic network models~\cite{,majdandzic2014,bottcher2017critical,bottcher2017failure}, then to active fluids~\cite{siebert2018critical,partridge2019critical,dittrich2020critical}. And we notice that our analysis for $T>0$ is different from  recent theoretical works~\cite{bottcher2017critical,bottcher2017failure}, in which the authors concluded a  DP-type of the transition by simply doing perturbation along the orthogonal directions in the $(a,~b)$ parameter space. {Therefore, more numerical evidences are required to settle this issue.}

\section{Conclusion and Discussion}
In conclusion, by using simulation and theoretical analysis, we systematically investigate the criticality of a reactive hard sphere system with an activation barrier. We find that increasing the activation barrier effectively delays the dynamic transition but also increases the transition cooperativity  which sharpens the transition.  {\color {black} The cooperativity enhancement comes from the interplay between activation barrier and the inertia of particles which make it possible for previous activation to facilitate the consecutive ones.}  This mechanism of barrier-controlled criticality has a direct implication for exothermic chemical reactions, where it was suggested that increasing the activation energy can make the  reactions change from slow combustion to thermal explosion~\cite{boddington1979t,lacey1983critical}.
Similar mechanism { may} also exist in nuclear chain reactions, which is highly relevant with the nuclear criticality safety~\cite{lamarsh2001introduction}.
Moreover, our finding {is related} with the dynamic phase transition in random organization systems. For examples, it was reported experimentally that for high density colloidal suspensions under oscillatory shearing, the geometrical protection from neighbouring particles suppresses or cancels the activated displacement, which sharpens the transition~\cite{j2014geometric}. Similar discontinuous dynamic phase transitions have also been observed in driven amorphous solids~\cite{nagasawa2019,das2020unified}, the glass transition systems~\cite{hedges2009dynamic} and high-density active matter systems~\cite{tjhung2017,lei2019}. In these cases, particles need to cross  energy barriers or cages set by their neighbors to be activated. Therefore, our finding about the activation barrier on the criticality of  a miminal reactive hard sphere model,  can not only help understand the criticality of reactive particle systems, e.g., chemical reactions, but also shed lights in the  dynamic behaviour of  amorphous materials, active matter, as well as  the spread of epidemic, knowledge and neural signals~\cite{dodds2004universal, g2002spiking, pastor2015epidemic}.

\subparagraph{Acknowledgments:} This work has been supported in part by the Singapore Ministry of Education through the Academic Research Fund MOE2019-T2-2-010 and RG104/17 (S), by Nanyang Technological University Start-Up Grant (NTU-SUG: M4081781.120), by the Advanced Manufacturing and Engineering Young Individual Research Grant (A1784C0018) and by the Science and Engineering Research Council of Agency for Science, Technology and Research Singapore, { by the National Natural Science Foundation of China under Grant No. 11905001}. We thank NSCC for granting computational resources.

\section{Appendix A: Solution of cubic dynamic equation}
Considering the general cubic dynamic equation,
\begin{eqnarray}\label{MF_dy}
\frac{\partial \tilde{T}_{k}}{\partial \tilde{t} } = a \tilde{T}_{k} -b \tilde{T}_{k}^2 -c \tilde{T}_{k}^3 + h
\end{eqnarray}

 In the case of zero conjugated field or thermal noise ($h=0$), Eq.(\ref{MF_dy}) have three steady states with $\frac{\partial  \tilde{T}_{k}}{\partial t}=0$ ,
\begin{eqnarray}
\tilde{T}_{k} = 0; ~~~~~~ \tilde{T}_{k} = -\frac{b}{2c} \pm \sqrt{ \frac{a}{c} + \left( \frac{b}{2c} \right)^2  }   
\end{eqnarray}
One can prove that when $b>0$, the Eq.(\ref{MF_dy}) shows a transcritical bifurcation and there is one fixed point, i.e., 
\begin{eqnarray}
\tilde{T}_{k} = -\frac{b}{2c} + \sqrt{ \frac{a}{c} + \left( \frac{b}{2c} \right)^2 }
\end{eqnarray}
This corresponds to a second-order phase transition with critical scaling  $ \tilde{T}_{k}\sim a^{\beta_{\rm MF}}$ and critical exponent $\beta_{\rm MF}=1$ as $a \rightarrow 0$. For $b<0$ the Eq.(\ref{MF_dy}) has two fixed points, i.e.,
\begin{eqnarray}
\tilde{T}_{k} = 0; ~~~~~~ \tilde{T}_{k} = -\frac{b}{2c} + \sqrt{ \frac{a}{c} + \left( \frac{b}{2c} \right)^2  }   
\end{eqnarray}
which corresponds to a bistablity or a dis-continuous dynamic phase transition. Importantly, $b=0$ is the tricritical point  separating the continuous and dis-continuous phase transitions. At this point, the system exhibits supercritical pitchfork bifurcation and the mean-field tricritical scaling behavior changes into
\begin{eqnarray}
\tilde{T}_{k} =  \sqrt{ \frac{a}{c} } \sim a^{\beta^t_{\rm MF}}
\end{eqnarray}
with the mean-field tricrtical exponent $\beta^t_{\rm MF}=1/2$ different from the $b>0$ case. The upper critical dimension also decreases from 4  in the previous one to 3 at this tricritcal point~\cite{lubeck2006tricritical}. For non-zero conjugated field ($h>0$), the boundary of the biastablity region is determined by the following equation,
\begin{eqnarray}
\left( \frac{h}{2c} -\frac{ab}{6c^2} -\frac{b^3}{27c^3}   \right)^2 = \left(\frac{a}{3c} + \frac{b^2}{9c^2} \right)^3   ~~  (b<0, ~a<0).~~
\end{eqnarray}
The theory also predicts a critical line as function of $h>0$, i.e.,
\begin{eqnarray}
 b_c= -3\left(\frac{h}{c}\right)^{1/3};~~
 a_c= -3\left(\frac{h}{c}\right)^{2/3};
 ~~\tilde{T}_{k,c}= \left(\frac{h}{c}\right)^{1/3}.~~
\end{eqnarray}
To obtain the mean-field critical exponent on this line, we do the perturbation of $b$ around an arbitrary critical point on this line, i.e.,
\begin{eqnarray}\label{MF_pertur}
b &=& b_c+\Delta b \\
\tilde{T}_{k} &=& \tilde{T}_{k,c} + \Delta \tilde{T}_{k}.
\end{eqnarray}
Based on Eq.~(\ref{MF_dy}), we have 
\begin{eqnarray}\label{MF_pertur2}
\frac{\partial \tilde{T}_{k} }{\partial \tilde{t}} &=& -c\left[\tilde{T}_{k,c} + \Delta \tilde{T}_{k} -\left(\frac{h}{c}\right)^{1/3} \right]^3  -  \Delta b(\tilde{T}_{k,c} + \Delta \tilde{T}_{k} )^2  \nonumber \\
 &\simeq &   -c \Delta \tilde{T}_{k}^3  -  \Delta b\tilde{T}_{k,c}^{2}=0
\end{eqnarray}
which leads to the response of the order parameter
\begin{eqnarray}
\Delta \tilde{T}_{k} \simeq -\left(\frac{\Delta b}{c}\right)^{1/3}{\tilde{T}_{k,c}^{2/3}}
\end{eqnarray}
A similar perturbation of $a$ around  the critical point gives
\begin{eqnarray}
\Delta \tilde{T}_{k} \simeq  \left(\frac{\Delta a}{c}\right)^{1/3}{\tilde{T}_{k,c}^{1/3}}
\end{eqnarray}
This  leads to mean-field critical exponent $\sigma_{\rm {MF}}^{-1} = 1/3$, the same as that in the mean-field Ising model. To obtain crtical exponent $\beta_{\rm MF}$, we do perturbation along the path $\Delta a = \Delta b \tilde{T}_{k}^*$. We have
\begin{eqnarray}\label{MF_pertur4}
\frac{\partial \Delta \tilde{T}_{k}}{\partial \tilde{t}} &=& -c \Delta \tilde{T}_{k}^3  + \Delta a{(\tilde{T}_{k,c}+\Delta \tilde{T}_{k})}-   \Delta b{(\tilde{T}_{k,c}+\Delta \tilde{T}_{k})}^{2}  \nonumber  \\
&=& -c \Delta \tilde{T}_{k}^3  -  \tilde{T}_{k,c} \Delta b\Delta \tilde{T}_{k} - \Delta b \Delta \tilde{T}_{k}^2  \nonumber  \\
&\simeq &-c \Delta \tilde{T}_{k} ( \Delta \tilde{T}_{k}^2 + \tilde{T}_{k,c} \Delta b/c)  =0 
\end{eqnarray}
Therefore, if $\Delta b <0$, the system  has two stable fixed points with emergent ${\mathbb Z}_2$ symmetry,
\begin{eqnarray}
\Delta \tilde{T}_{k} \simeq  \pm \sqrt{-\Delta b \tilde{T}_{k,c}/c }.
\end{eqnarray}

\bibliographystyle{nature}
\bibliography{reference,reference2}

\end{document}